\documentclass[twocolumn,showpacs]{revtex4}

\usepackage{graphicx}
\usepackage{dcolumn}
\usepackage{amsmath}
\begin{document}  
\title{Thermal activation between Landau levels in the organic
superconductor
$\beta''$-(BEDT-TTF)$_{2}$SF$_{5}$CH$_{2}$CF$_{2}$SO$_{3}$}

\author{M.-S.~Nam$^1$, A.~Ardavan$^1$, J.A.~Symington$^1$,
J.~Singleton$^{1,2}$, N.~Harrison$^2$, C.H.~Mielke$^2$,
J.A.~Schlueter$^3$, R.W.~Winter$^4$ and G.L.~Gard$^4$}
\affiliation{$^1$Department of Physics, University of Oxford, Clarendon
Laboratory, Parks Road, Oxford OX1 3PU, United Kingdom.\\
$^2$National High Magnetic Field Laboratory, LANL, MS-E536, Los
Alamos, New Mexico 87545, USA\\
$^3$Chemistry and Materials Science Division,
Argonne National Laboratory, Argonne, Illinois 60439\\
$^4$Department of Chemistry, Portland State University, Portland, Oregon 97207, USA}

\begin{abstract} 
We show that Shubnikov-de Haas oscillations
in the interlayer resistivity of the
organic superconductor  $\beta''$-(BEDT-TTF)$_{2}$SF$_{5}$
CH$_{2}$CF$_{2}$SO$_{3}$ become very pronounced in
magnetic fields $\sim$~60~T.
The conductivity minima exhibit thermally-activated behaviour
that can be explained simply by the presence of a Landau gap,
with the quasi-one-dimensional Fermi surface sheets contributing
negligibly to the conductivity. This observation, together with
complete suppression of chemical potential oscillations,
is consistent with an incommensurate nesting instability of the
quasi-one-dimensional sheets.
\end{abstract}


\maketitle

The quantizing effect of a magnetic field on a
charge-carrier system is well known~\cite{ashcroft-mermin}. In
metals, this leads to oscillations of the free energy
and quasiparticle density of states as Landau levels cross the
Fermi energy~\cite{Shoenberg}. The
effect is very pronounced in quasi-two-dimensional (Q2D)
metals containing Fermi surfaces (FSs) that are approximately
cylindrical~\cite{john-review}. Recently there has also been interest in
analogous effects in quasi-one-dimensional (Q1D)
FS sections, which can lead to magnetic-field-induced
quantization~\cite{yakovenko} and
localization~\cite{Dupuis}.

In this paper we describe the magnetoresistance of the
organic superconductor
$\beta''$-(BEDT-TTF)$_{2}$SF$_{5}$CH$_{2}$CF$_{2}$SO$_{3}$
($T_{\rm c}\approx$5.4~K~\cite{geiser}).
Bandstructure calculations suggest that this material
possesses a FS comprising a Q2D
cylinder and a pair of Q1D sheets~\cite{geiser}.
However, Shubnikov-de Haas (SdH) and
de Haas-van Alphen (dHvA) measurements reveal
that the Q2D cylinder has only one third the expected
cross-section~\cite{JW1,JW2,Su}.
Angle-dependent magnetoresistance oscillation (AMRO)~\cite{JW2,am2} and
millimetre-wave magnetoconductivity experiments~\cite{edwards}
show that the cross-section of this Q2D pocket resembles
an elongated diamond. The same experimental techniques
find no evidence for the presence of Q1D
Fermi sheets at low temperatures~\cite{JW2,am2,edwards},
unlike the situation in other Q2D organic metals~\cite{john-review}.
By contrast, in order to explain the observation of
a fixed chemical potential $\mu$ in the dHvA effect~\cite{JW1},
Wosnitza {\it et al.} proposed Q1D
states which have an enormous
density of states,
exceeding the estimates from bandstructure
calculations
by at least an order of magnitude~\cite{JW1}.
Moreover, using a simple formula
for the background magnetoresistance, Wosnitza {\it et al.}
suggested that the Q1D
sheets become 
localised in a magnetic field \cite{JW3}.
In the present paper, we show that magnetoresistance
data suggest a
much simpler explanation.
The thermally activated behaviour of the data at
integer
Landau level filling factors is explained entirely in terms of
a Landau gap. 
Moreover, the failure of the Q1D sheets to
contribute to the conductivity together with their ability to fix
$\mu$ is explained by their nesting to
form an incommensurate density-wave
ground state. This mechanism is supported by the
temperature dependence of the resistivity at
$B=$~0.

Single crystals ($\sim 1.0 \times 0.5\times 0.2$~mm$^{3}$)
of $\beta''$-(BEDT-TTF)$_{2}$SF$_{5}$CH$_{2}$CF$_{2}$SO$_{3}$
were prepared using standard electrochemical
techniques~\cite{geiser}. Contacts were applied
using 12.5~$\mu$m Pt wires and graphite paint, in a
configuration which gives the interplane magnetoresistance,
$\rho_{zz}$~\cite{john-review}.
Magnetic fields $B$ were provided by
60~T pulsed magnets at the National High Magnetic Field Laboratory
(NHMFL), Los Alamos; two magnet pulses
were required for each data set, the {\it dc}
sample current of $5~\mu$A
being reversed between pulses
to remove contributions from thermoelectric
and inductive voltages. Temperatures $T$
between 0.5~K and 4.2~K were obtained using
a $^3$He cryostat. Sample heating was not found to be a
problem. This was checked by ensuring that the measured
$\rho_{zz}$ was independent of ${\rm d}B/{\rm d}t$ and by comparing
data with those taken in quasistatic fields of up to
$\sim$~33~T at NHMFL Tallahassee.

\begin{figure}[tbp]
   \centering
\includegraphics[height=12cm]{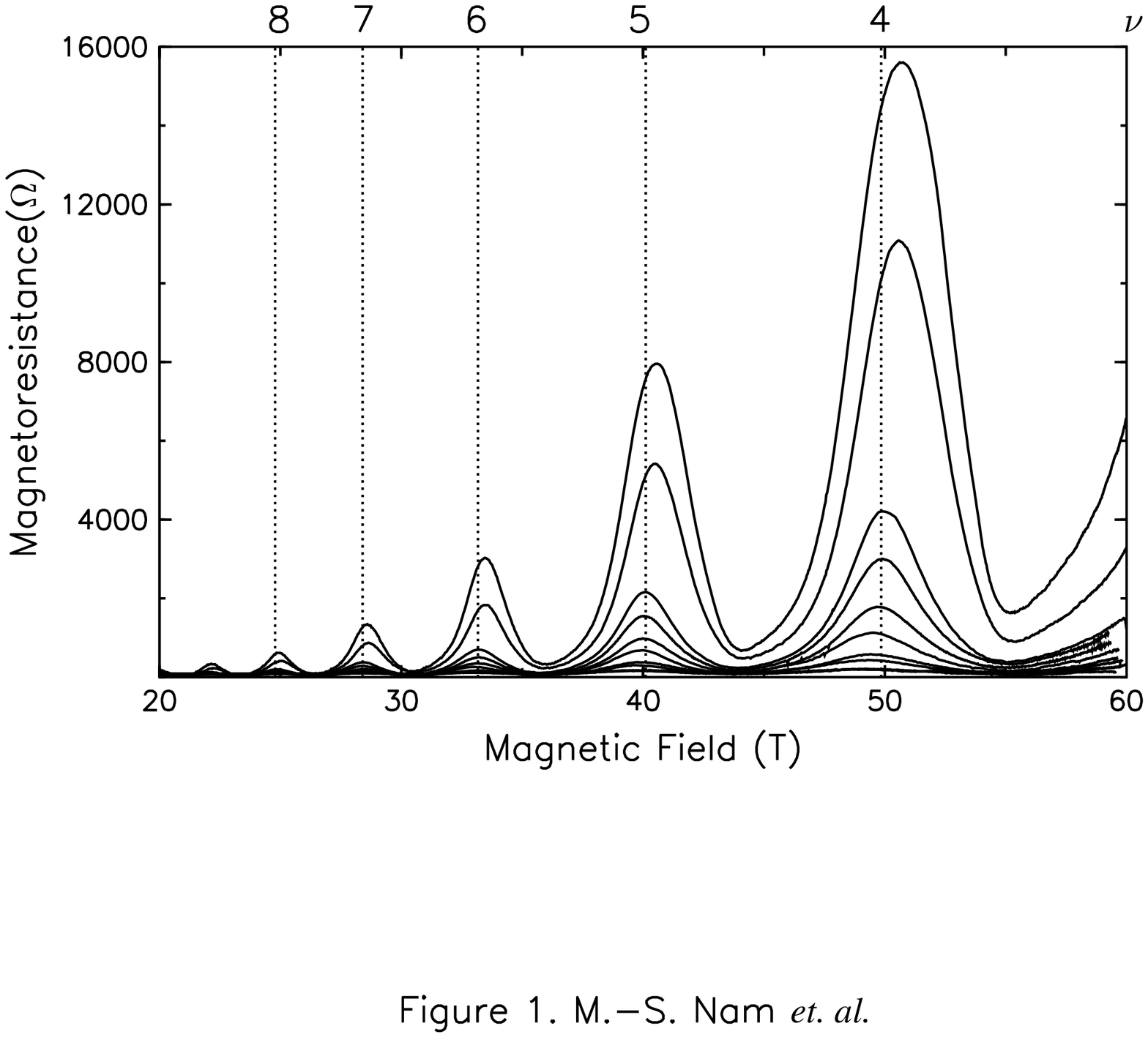}
\caption{The temperature dependent magnetoresistance in
$\beta''$-(BEDT-TTF)$_{2}$SF$_{5}$CH$_{2}$CF$_{2}$SO$_{3}$ (from the
top, 0.59, 0.94, 1.48, 1.58, 1.91, 2.18, 2.68, 3.03, 3.38, 3.80, and
4.00~K). The dotted lines and numbers indicate integer Landau-level
filling factors $\nu=F/B$.}
\label{sdh}  
\end{figure}

Fig.~\ref{sdh} shows the $T$-dependent
resistivity $\rho_{zz}$; $B$ was applied
perpendicular to the Q2D planes. Pronounced SdH oscillations
are visible, with a frequency $F=196 \pm 3$~T in
agreement with earlier data~\cite{JW1,JW2,Su,edwards}.
At high fields, $\rho_{zz}$ becomes very large close
to integer Landau-level filling-factor, $\nu=F/B$.

In three-dimensional (3D) metals exhibiting the SdH effect,
$\Delta\sigma/\sigma$, the ratio of the oscillatory
part of the conductivity to the background conductivity
(originating from the rest of the FS)
is proportional to $B^2{\rm d}M/{\rm d}B$, where
$M$ is the magnetisation~\cite{Shoenberg}. This is valid in 3D metals
because $\Delta \sigma$ is only a very weak
perturbation of $\sigma$. However,
this formula is no longer applicable
in a Q2D system in which $\Delta\sigma \gg \sigma$~\cite{neil1},
as is clearly the case in Fig.~\ref{sdh}. For such data,
the value of $\sigma$ that one extracts by comparing
$\Delta\sigma/\sigma$ with $B^2{\rm d}M/{\rm d}B$ (as done in Ref.~\cite{JW3})
has no real
physical meaning~\cite{john-review,neil1}.
In Ref.~\cite{neil1}, it was shown that in such situations,
$\rho_{zz}$ behaves in an insulating fashion whenever
$\mu$ resides in a Landau gap, which occurs when
$\nu=F/B$ is an integer. It is for this reason
that the large peaks in $\rho_{zz}$ occur close to integer $\nu$
in Fig.~\ref{sdh}; if one
incorrectly extracted $\sigma$ from the SdH data~\cite{JW3}
this effect might well lead one to suspect that
a ``field-induced insulator''~\cite{JW3} had occurred.

The effective mass $m^{\ast}$ and Dingle temperature
$T_{\rm D}$~\cite{Shoenberg,Springford} were deduced from SdH
oscillations between 2 and 6~T, measured in the quasistatic
magnet. This corresponds to $59 \leq \nu \leq 33$,
where the application of 2D
Lifshitz-Kosevich (LK) analysis is
able to extract reliable values of $m^{\ast}$~\cite{Shoenberg,john-review}.  The
values deduced are $m^{\ast}=1.96 \pm 0.05~m_{\rm e}$ and
$T_{\rm D}=0.82$~K. Using
$T_{\rm D} = \hbar /2 \pi k_{\rm B} \tau$~\cite{Springford},
the scattering time $\tau$ is
$1.22 \times 10^{-12}$~s, leading to a half-width
broadening of the Landau levels~\cite{Shoenberg,Springford}
$\delta E =\hbar \tau^{-1}= 0.44$~meV.
By comparison, in the absence of
broadening, the energy
gap between adjacent Landau levels is $\hbar
\omega_{c}=\hbar eB/m^{\ast}$;
using $m^{\ast}$ from the
LK analysis, $\hbar \omega_{\rm c}= 2.94$~meV at
49.85~T ($\nu = 4$).

\begin{figure}[tbp]
   \centering
\includegraphics[height=12cm]{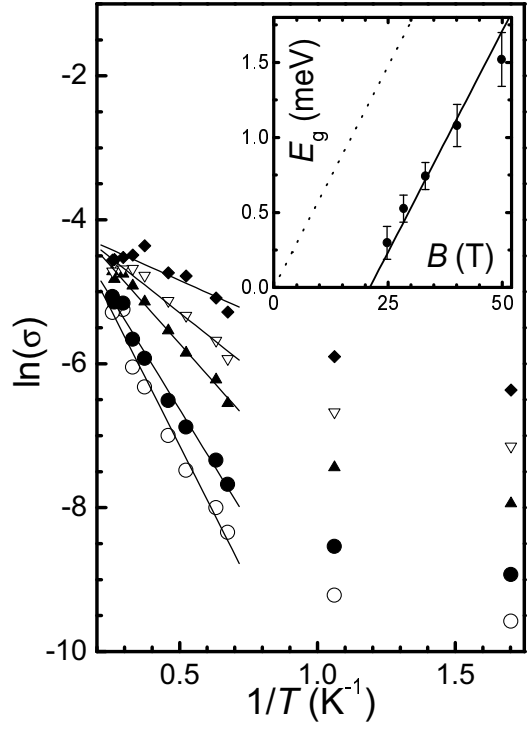}
\caption{$\sigma_{zz}$ against
$1/T$ at the magnetoresistance peaks
close to integer filling factors; filled diamond $\nu =
8$, open triangle $\nu = 7$, filled triangle $\nu = 6$, filled circle
$\nu = 5$, and open circle $\nu = 4$.
The lines are fits used to extract
$\Delta$
(see Eqn.~\ref{bollocks}).
The inset shows the magnetic
field dependence of $E_{\rm g}=2\Delta$.
The dotted line shows $E=\hbar \omega_{\rm c}=\hbar eB/m^{\ast}$
for comparison
and the solid line shows $E=\hbar \omega_{\rm c}-E_{\rm O}$,
where $E_{\rm O}$ is a constant offset energy (see text).
}
\label{activation} 
\end{figure}

At the maxima close to integer $\nu$, $\rho_{zz}$
increases exponentially with decreasing
$T$. Fig.~\ref{activation} shows
$\log\sigma_{zz} = \log(1/\rho_{zz})$ at these points versus $1/T$.
For $1.4<T<3.9$~K
the data show clear
thermal activation behaviour,
\begin{equation}
\sigma_{zz} \propto {\rm e}^{-\Delta/k_{\rm B}T};
\label{bollocks}
\end{equation}
at lower $T$, the drop in conductivity
saturates.
The slope of each line in Fig.~\ref{activation}
gives $\Delta$, and the inset shows $E_{\rm g}=2\Delta$
versus $B$ (the reason for plotting $2\Delta$ will become clear below);
$E_{\rm g}$ increases linearly with $B$, suggesting that the
gap is related to the Landau quantization.

\begin{figure}[tbp]
   \centering
\includegraphics[height=12cm]{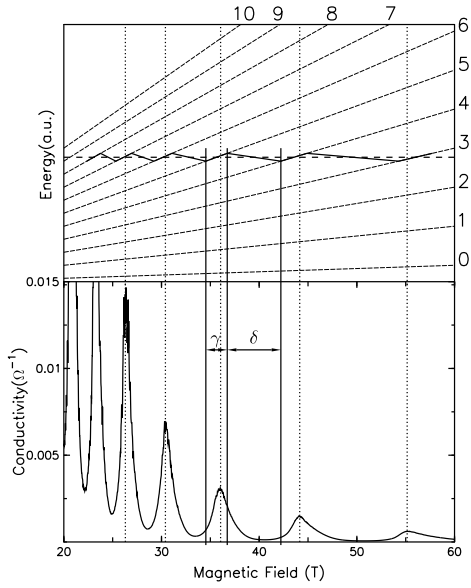}
\caption{The upper figure shows the Landau fan diagram with a
schematic view of the motion of the chemical potential $\mu$ where $T$ and
$\tau^{-1}$ are zero. The solid line indicates the motion
of $\mu$ when it is alternately pinned in Q1D and Q2D states.
The dashed line shows $\mu$ held constant by
{\it e.g.} an incommensurate density wave.
The lower part
shows the magnetoconductivity oscillation at 500\,mK; the conductivity peaks
when $\mu$ is between Landau levels.
} \label{chemical}
\end{figure}

The possible explanations for this behaviour
involve the movement of the Landau levels
in
$\beta''$-(BEDT-TTF)$_{2}$SF$_{5}$CH$_{2}$CF$_{2}$SO$_{3}$
with respect to $\mu$ as a function of $B$.
In a perfect ($\tau=\infty$) metal comprising a single
Q2D Fermi surface at $T=0$, $\mu$ is always
pinned to a Landau level in a magnetic field~\cite{Shoenberg}.
With increasing
$B$, $\mu$ moves up in a particular
Landau level until the degeneracy of the levels
below has increased sufficently for them to accommodate
all the quasiparticles. At this point,
$\mu$ drops
discontinuously into the Landau level below~\cite{Shoenberg}.
The presence of Q1D Fermi sheets, which give a continuous
dispersion in a magnetic field~\cite{yakovenko}, modifies
this behaviour~\cite{john-review,neil1}. In this case, shown in Fig.~\ref{chemical},
$\mu$ (solid line) is
alternately pinned to a Landau level or to the Q1D density of states as the
field increases~\cite{neil1}.
While $\mu$
is  pinned to a Landau level (region $\gamma$ in
Fig.~\ref{chemical}) the system acts as a
Q2D metal. In the region labelled $\delta$ (Fig.~\ref{chemical}),
$\mu$ is between two adjacent Landau levels, and here
only the Q1D dispersion contributes to the conduction.
It is in the latter regions that
we observe insulating behaviour,
implying that the Q1D Fermi sheets do not
contribute to $\sigma_{zz}$ at low $T$
and high $B$.
Q1D quasiparticles are
expected to undergo magnetic-field-induced
localization~\cite{Dupuis} at high fields.
However, this cannot explain the data.
Localisation will either lead to a broad band
of immobile states
or the collapse of the Q1D states onto
states of discrete energy~\cite{mott}.
The former would lead to
conduction at integer $\nu$ in which
quasiparticles were thermally
excited from the band of immobile
states to the
Landau levels; this would result
in conductivity obeying a power law.
The latter scenario removes the mechanism for
pinning $\mu$ between Landau levels, and $\mu$ would instead
drop discontinuously between the levels.

\begin{figure}[tbp]
   \centering
\includegraphics[height=12cm]{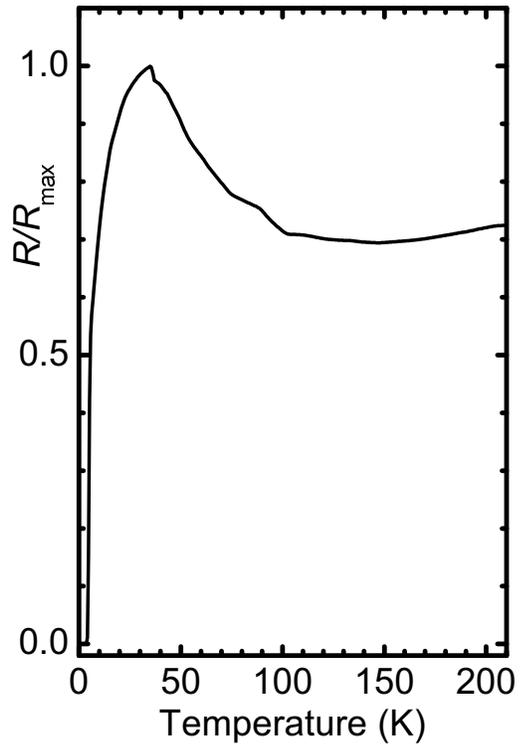}
\caption{The temperature dependence of the interlayer
resistance ($\propto \rho_{zz}$) normalised to
its maximum value, measured with a current of
$0.5~\mu$A. With currents any higher than this, $R$ exhibits
bistability and hysteresis between 35~K and 140~K.
} \label{tdep}
\end{figure}

A clue to the cause of the activated behaviour
is given by Fig.~\ref{tdep}, which shows the $T$-dependence of
$\rho_{zz}$ at $B=0$; $\rho_{zz}$ initally decreases with decreasing $T$
until $T \approx 140$~K, where there is a minimum.
Thereafter, $\rho_{zz}$ increases with decreasing $T$
until about 35~K, from where it drops to the superconducting transition
at $T\approx 5$~K. In the region $35<T<140$~K, the measured
$\rho_{zz}$ depends on the current used in the experiment,
and can exhibit jumps, bistability and hysteresis under certain
conditions of bias.
The latter behaviour, and the minimum at 140~K are
typical of organic density-wave (DW) systems, {\it e.g.}
(BEDT-TTF)$_3$Cl$_2$.2H$_2$O~\cite{lub}
(however, in $\beta''$-(BEDT-TTF)$_{2}$SF$_{5}$CH$_{2}$CF$_{2}$SO$_{3}$
the behaviour is somewhat less extreme than that of
(BEDT-TTF)$_3$Cl$_2$.2H$_2$O~\cite{lub}, presumably because the
Q2D pockets of the FS survive the transition).
We therefore propose that the
Q1D Fermi sheets in
$\beta''$-(BEDT-TTF)$_{2}$SF$_{5}$CH$_{2}$CF$_{2}$SO$_{3}$
nest to form a DW state at $T \approx 140$~K.
A DW transition is known
to occur at much lower $T$ in the isostructural salt
$\beta''$-(BEDT-TTF)$_{2}$AuBr$_{2}$~\cite{house};
however, the nesting is
imperfect, leading to a low-temperature FS comprising a number of
small Q2D pockets in addition to an elongated pocket similar to that
found in 
$\beta''$-(BEDT-TTF)$_{2}$SF$_{5}$CH$_{2}$CF$_{2}$SO$_{3}$~\cite{house}. In
$\beta''$-(BEDT-TTF)$_{2}$SF$_{5}$CH$_{2}$CF$_{2}$SO$_{3}$, the lack of such
additional pockets and the higher ordering $T$
suggest that the nesting
is more efficient. The nesting would explain the absence of any signature
of the Q1D sheets in the AMRO~\cite{JW2,am2} and millimetre-wave
data~\cite{edwards}.

It is known that incommensurate DWs
adjust their nesting vectors ${\bf Q}$ in a magnetic field in
order 
to minimise the total free energy~\cite{chaikin}.
In the Bechgaard salts, this is lowest whenever $\mu$ is situated in a
Landau gap of the pocket created by imperfect nesting of the Q1D
sheets~\cite{chaikin}. This
mechanism will not, however, operate if the nesting is
perfect~\cite{Harrison}. On the other hand, it was recently shown that the
oscillations of $\mu$ originating from
Landau quantisation of additional Q2D
FS sections also affect the free energy~\cite{Harrison}.
Two qualitatively different
types
of behaviour can result~\cite{Harrison}; in the
case of a commensurate DW, the DW
can become unstable to the
oscillations
of $\mu$. In the incommensurate case, however, ${\bf
Q}$ is free to adjust itself to prevent
the variation of $\mu$ and thereby
suppress the positive $\mu^2$ term in the free energy. This mechanism
becomes
especially effective if the nesting is perfect so that
there is no residual pocket created by the Q1D sheets. In effect,
by adjusting {\bf Q}, a DW is able
to compensate for the oscillations in the filling of the Q2D
pocket (which result from the fact that $\mu$ is constant),
thereby functioning in the same way as an infinite charge reservoir~\cite{Harrison}.
We believe that this is happening in the present case.

If $\mu$ remains constant,
and the Landau levels sweep through it (as shown
by the dashed line in Figure~\ref{chemical}),
there will be no additional quasiparticle states between the
Landau levels; the only means of conduction
at integer $\nu$ is therefore thermal excitation
of quasiparticles from the full Landau level below $\mu$ (leaving behind
holes) into the empty level above it.
If the DW gap is much larger than the Landau gap
(as suggested by the transition temperature), the Q1D
sheets will only contribute to the conductivity at much higher $T$.
The situation is analogous to an intrinsic semiconductor;
the conductivity is proportional to the number $n$ of quasiparticles
excited, which in turn is given by the Law of
Mass-Action~\cite{ashcroft-mermin},
$\sigma_{zz} \propto n \propto {\rm e}^{-\frac{E_{\rm g}}{2 k_{\rm B}T}}$,
where $E_{\rm g}$ is the energy gap between the filled and empty Landau
level.
Comparing this with Eqn~\ref{bollocks}, we make the
identification $2\Delta \equiv E_{\rm g}$.

The inset to Fig.~\ref{activation} shows $E_{\rm g}$ versus $B$ plotted with
the function $E=\hbar \omega_{\rm c}-E_{\rm O}$ (solid line);
with $E_{\rm O}$ set to $1.23$~meV,
the experimental values of $E_{\rm g}$
all lie close to this line, strongly supporting
our proposal that $E_{\rm g}$
is related to the gap between the
Landau-level centres, $\hbar \omega_{\rm c}=\hbar eB/m^{\ast}$.
Landau-level broadening will cause a
reduction of the effective energy gap,
and it is this reduction that we identify with
$E_{\rm O}$;
$E_{\rm O}$ is about three times the Landau-level
half-width of 0.44~meV deduced from $T_{\rm D}$.
In a Lorentzian Landau-level density of states~\cite{Shoenberg,neil1},
the tails of the levels
extend beyond $\hbar \tau^{-1}$ from the centre; even when
$\mu$ lies between two Landau levels, there will be a
small number of current-carrying quasiparticles. This is probably
why the
conductivity saturates at low $T$; the carriers in
the tails of the Landau levels play the role of extrinsic carriers in
a semiconductor system~\cite{ashcroft-mermin}.

Finally, we note that the width of the ``tail'' of the
Fermi-Dirac distribution function~\cite{Shoenberg}
($\sim 6k_{\rm B} T$) is still comparable to
the Landau-level width $\hbar / \tau \approx 0.44$~meV,
even at $T \approx 590$~mK ($6 k_{\rm B}T \sim 0.3$~meV).
This causes the resistivity at half-integer
filling factors to increase with decreasing $T$~\cite{druist},
but to a much lesser extent
than at integer filling factors (Fig.~1).
This effect is well-documented in the
interlayer magnetoresistance of semiconductor
superlattices (see e.g. \cite{druist})
and is {\it not} associated with field-induced localisation~\cite{JW3}.

In summary, we have observed thermally-activated
conductivity at integer Landau-level filling factors
in the Q2D
organic metal $\beta''$-(BEDT-TTF)$_{2}$SF$_{5}$CH$_{2}$CF$_{2}$SO$_{3}$.
To our knowledge, this is the first identification of such a mechanism in
a metallic system.
The activation energies deduced from the conductivity are in good
agreement with the Landau-level spacings once broadening is taken into
account suggesting that the recent identification of
a ``field-induced insulator''~\cite{JW3} in this material is
incorrect.
To account for this behaviour and for magnetisation
and resistivity data, we propose that the Q1D sheets of the Fermi surface of
$\beta''$-(BEDT-TTF)$_{2}$SF$_{5}$CH$_{2}$CF$_{2}$SO$_{3}$ are nested.
The temperature dependence of the resistivity suggests that this
occurs at $T \approx 140$~K.

This work is supported by EPSRC (UK).
NHMFL is supported by the
US Department of Energy (DoE), the National
Science Foundation and the State of Florida.
Work at Argonne is sponsored
by the DoE, Office of Basic Energy Sciences,
Division of Materials Science under contract number
W-31-109-ENG-38.
We thank James Brooks
for stimulating discussions.

\clearpage

\end{document}